\documentclass[pdflatex,sn-mathphys-num]{sn-jnl}

\usepackage{graphicx}%
\usepackage{multirow}%
\usepackage{amsmath,amssymb,amsfonts}%
\usepackage{amsthm}%
\usepackage{mathrsfs}%
\usepackage[title]{appendix}%
\usepackage{xcolor}%
\usepackage{textcomp}%
\usepackage{manyfoot}%
\usepackage{booktabs}%
\usepackage{algorithm}%
\usepackage{algorithmicx}%
\usepackage{algpseudocode}%
\usepackage{listings}%
\usepackage{booktabs}
\usepackage{overpic}
\usepackage{subcaption}

\theoremstyle{thmstyleone}%
\theoremstyle{thmstyletwo}%

\theoremstyle{thmstylethree}%

\begin{document}


\title[Well-being and career instability across genders in the Spanish Astronomical Society]{Well-being and career instability across genders in the Spanish Astronomical Society}


\author*[1,2]{\fnm{Maritza A.} \sur{Lara-López}}\email{maritzal@ucm.es}

\author[3]{\fnm{Isabel} \sur{Rebollido}}

\author[4]{\fnm{Alba} \sur{Vidal-Garc\'ia}}

\author[5,3]{\fnm{Alicia} \sur{Rouco Escorial}}

\author[6,7,8]{\fnm{Sara R.} \sur{Berlanas}}

\author[6]{\fnm{Ismael} \sur{Garc\'ia-Bernete}}

\author[9]{\fnm{Beatriz} \sur{Ag\'is-Gonz\'alez}}

\author[3]{\fnm{Marina} \sur{Rodr\'iguez-Baras}}

\author[10]{\fnm{Naiara} \sur{Barrado-Izagirre}}

\author[11]{\fnm{Irene} \sur{Pintos Castro}}

\author[12]{\fnm{Nataly} \sur{Ospina}}

 \author[13,14]{\fnm{Silvia} \sur{Bonoli}}

\affil*[1]{\orgdiv{Departamento de Física de la Tierra y Astrofísica}, \orgname{Fac. de C.C. Físicas, Universidad Complutense de Madrid}, \orgaddress{\postcode{28040}, \city{Madrid}, \country{Spain}}}

\affil*[2]{\orgdiv{Instituto de Física de Partículas y del Cosmos, IPARCOS}, \orgname{Fac. de C.C. Físicas, Universidad Complutense de Madrid}, \orgaddress{\postcode{28040}, \city{Madrid}, \country{Spain}}}
    
\affil[3]{\orgname{European Space Agency (ESA), European Space Astronomy Centre (ESAC)}, \orgaddress{\street{Camino Bajo del Castillo s/n}, \city{Villanueva de la Ca\~nada, Madrid}, \postcode{28692},  \country{Spain}}}

\affil[4]{\orgname{Observatorio Astronómico Nacional}, \orgaddress{\street{C/ Alfonso XII 3}, \city{Madrid}, \postcode{28014}, \country{Spain}}}

\affil[5]{\orgname{Starion España SLU}, \orgaddress{\street{C/Chile 10, second floor, office 247}, \city{Las Rozas, Madrid}, \postcode{28290}, \country{Spain}}}

\affil[6]{\orgdiv{Centro de Astrobiolog\'ia (CAB)},\orgname{CSIC-INTA}, \orgaddress{\street{Camino Bajo del Castillo s/n}, \city{Villanueva de la Ca\~nada, Madrid}, \postcode{28692}, \country{Spain}}}

\affil[7]{\orgname{Dpto. de Astrofísica, Universidad de La Laguna}, \city{Tenerife}, \postcode{38205}, \country{Spain}}

\affil[8]{\orgname{Instituto de Astrofísica de Canarias}, \city{Tenerife}, \postcode{38200}, \country{Spain}}

 \affil[9]{\orgdiv{Institute of Astrophysics}, \orgname{FORTH}, \orgaddress{\street{N. Plastira 100, Vassilika Vouton}, \city{Heraklion}, \postcode{70013}, \country{Greece}}}

 \affil[10]{\orgdiv{Dpto. Física Aplicada, Escuela de Ingeniería de Bilbao}, \orgname{University of the Basque Country (UPV/EHU)}, \orgaddress{\street{Plaza Ingeniero Torres Quevedo 1}, \city{Bilbao},  \country{Spain}}}

 \affil[11]{\orgname{FlowReserve Labs S.L.}, \orgaddress{\city{Santiago de Compostela}, \postcode{15706}, \country{Spain}}}

 \affil[12]{\orgname{INFN Sezione di Bari}, \orgaddress{\street{Via Orabona 4},  \city{Bari}, \postcode{70126}, \country{Italy}}}

 \affil[13]{\orgdiv{Donostia International Physics Center}, \orgname{DIPC}, \orgaddress{\street{Paseo Manuel de Lardizabal 4}, \city{Donostia-San Sebastian},  \postcode{20018}, \country{Spain}}}

 \affil[14]{\orgdiv{Basque Foundation for Science}, \orgname{IKERBASQUE}, \city{Bilbao},  \postcode{48013}, \country{Spain}}







%
\abstract{We present the results of a comprehensive survey conducted among members of the Spanish Astronomical Society (Sociedad Española de Astronomía, SEA) to assess well-being, professional satisfaction, and family–work balance of researchers in astronomy. The survey addressed multiple aspects of professional life, including happiness, career stability, publication pressure, and access to childcare services during scientific meetings. Responses were examined across gender and career stages to identify trends and sources of dissatisfaction.



}

\maketitle

\section{Introduction}\label{sec1}

Amid growing international concern regarding the mental health crisis within academia, the Women and Astronomy Commission (Comisión Mujeres y Astronomía, CMyA, https://www.sea-astronomia.es/comision-mujer-y-astronomia) of the SEA (https://www.sea-astronomia.es/) took the initiative to design and distribute a well-being survey aimed at  researchers who are members of the SEA. The objective of this survey was to gain insights into the current state of mental health and well-being within our community, identify potential challenges, and highlight areas where institutional support may be most needed.

The survey consisted of 17 core questions, several of which included follow-up items to allow for more detailed responses. To ensure broad participation, the questionnaire was distributed through the SEA members mailing list. This outreach resulted in a total of 218  responses, corresponding to roughly 33\% of all SEA full members with a PhD, providing a valuable dataset that reflects a wide range of experiences and perspectives within the Spanish astronomical community.

Two main topics were covered in this survey. The first part focused on demographics, with questions on gender, pronouns, citizenship, year of PhD defense, career stage (ranging from postdoc, telescope operator, engineer, tenured professor/researcher, retired, left academia, or other), and fellowships obtained.

This section also included a set of well-being questions, such as an index of happiness, whether participants regretted pursuing a PhD, and whether they would recommend a student to do one. Additional questions explored the reasons for unhappiness (with multiple options provided), the most difficult part of their career, and whether they had considered leaving academia—and if so, why. The design of some survey questions was partly inspired by the survey conducted by the French Astronomical Society \cite{Boissier12}.


The second part of this survey concentrated on academic performance and productivity, including undergraduate marks and the number of publications, both during and after the PhD. This latter section will be the subject of a forthcoming publication.



\section{Demography}\label{sec2}

Currently, the Spanish Astronomical Society  comprises 655 full members holding a PhD (32.8\% women, 66.4\% men, and 0.8\% prefer not to answer/other), together with 316 predoctoral junior members (33.2\% women, 63.0\% men, and 3.8\% prefer not to answer/other). Membership of the SEA is open to professional astronomers and researchers working in astronomy and related fields, regardless of nationality or current country of employment. Given the nature of our survey, we focused on the PhD-holding membership.

From the 218 responses collected, 57.8\% correspond to men, 39.91\% to women, and 2.3\% to non-binary individuals or that preferred not to disclose their gender. 
Survey participants represent a diverse range of career stages and professional roles. For the purposes of the analyses presented in this paper, participants were grouped into three categories: Tenured Professors/Researchers, Postdoctoral Researchers, and Other. The latter category includes Engineers, Telescope Operators, Retired Members, Outreach Officers, Scientific or Technical Project Managers, Professors at Private Universities, and Unemployed Members. Of all participants, $\sim$23\% expressed the need of childcare services during SEA scientific meetings. Full demographic information is provided in Table~\ref{Demography}.

In terms of nationality, 86.2\% participants are Spanish, 10.6\% are from other European countries, and 3.2\% are from outside Europe.

\begin{table}[h]
\centering
\caption{Demography of gender and professional role of the survey participants. Numbers in parenthesis indicate participants in need of childcare.}
\begin{tabular}{lccccc}
\toprule
Gender & N\textsuperscript{\underline{o}} & \multicolumn{3}{c}{Professional role  } & In need of childcare\\
\cmidrule(lr){3-5}
       &            & Prof./Researcher & Postdoc & Other & \\ 
\midrule
\midrule

Women      & 87   & 36 (13) & 36 (12) & 15 (3)  & 28 \\
Male       & 126  & 69 (8)  & 43 (10) & 14 (5)  & 23 \\
Non-binary & 5     & 1    & 3   & 1    & 0 \\
\bottomrule
\end{tabular}
\label{Demography}
\end{table}

%
%


\section{Well-being}\label{Well-being}

Given the alarming concerns regarding mental health within the scientific community \cite{Trang2024, Webb2022, Webb2021}, it is essential to explore well-being alongside professional metrics. Since the degree of happiness is a complex and multifaceted issue, we sought to approximate an index of happiness rather than rely on a single question. To achieve this, we consulted with a social psychologist and framed our happiness-related question in the following way:

{\it We would like to understand the emotional well-being of the members of our society. Our profession can involve extensive travel, living abroad, and job instability, along with many enriching experiences and exciting science. Our level of motivation, expectations, and degree of professional satisfaction are constantly evolving, and our emotional well-being can be affected by both, our personal and professional life. Considering the balance between your research career and your personal well-being, are you happy?}

\begin{figure}[h]
\centering
\includegraphics[width=1.1\textwidth]{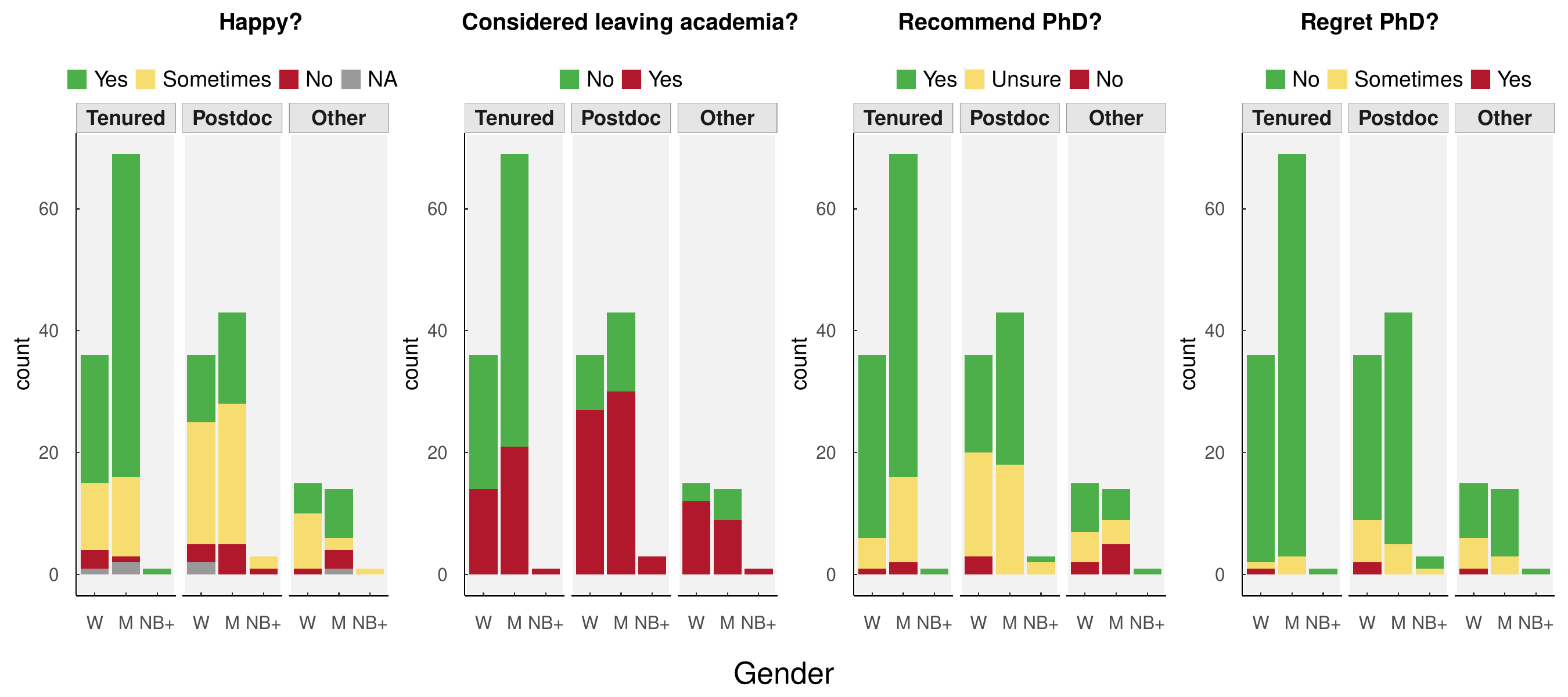}
\caption{Responses to the well-being questions by gender and professional stage. Each panel corresponds to the question indicated in the header. 
Responses are grouped by professional stage (Tenured  – Professors or Researchers –, Postdoctoral Researcher, and Other) and separated by gender (Women – W, Men – M, and Non-binary / Prefer not to say –  NB+). In the first panel, the “NA” category corresponds to Prefer not to answer.}\label{IndexHappy}
\end{figure}

In addition to the happiness question, we included several questions aimed at assessing the overall well-being of the Spanish astronomical community, these included: {\it Have you  considered leaving academia?, Would you recommend a student to pursue a PhD?, and Do you regret doing a PhD?}
The responses to these questions, cross-referenced with gender and professional stage of the participants, are presented in Fig.~\ref{IndexHappy}.

\begin{figure}
\centering

\includegraphics[width=1.1 \textwidth]{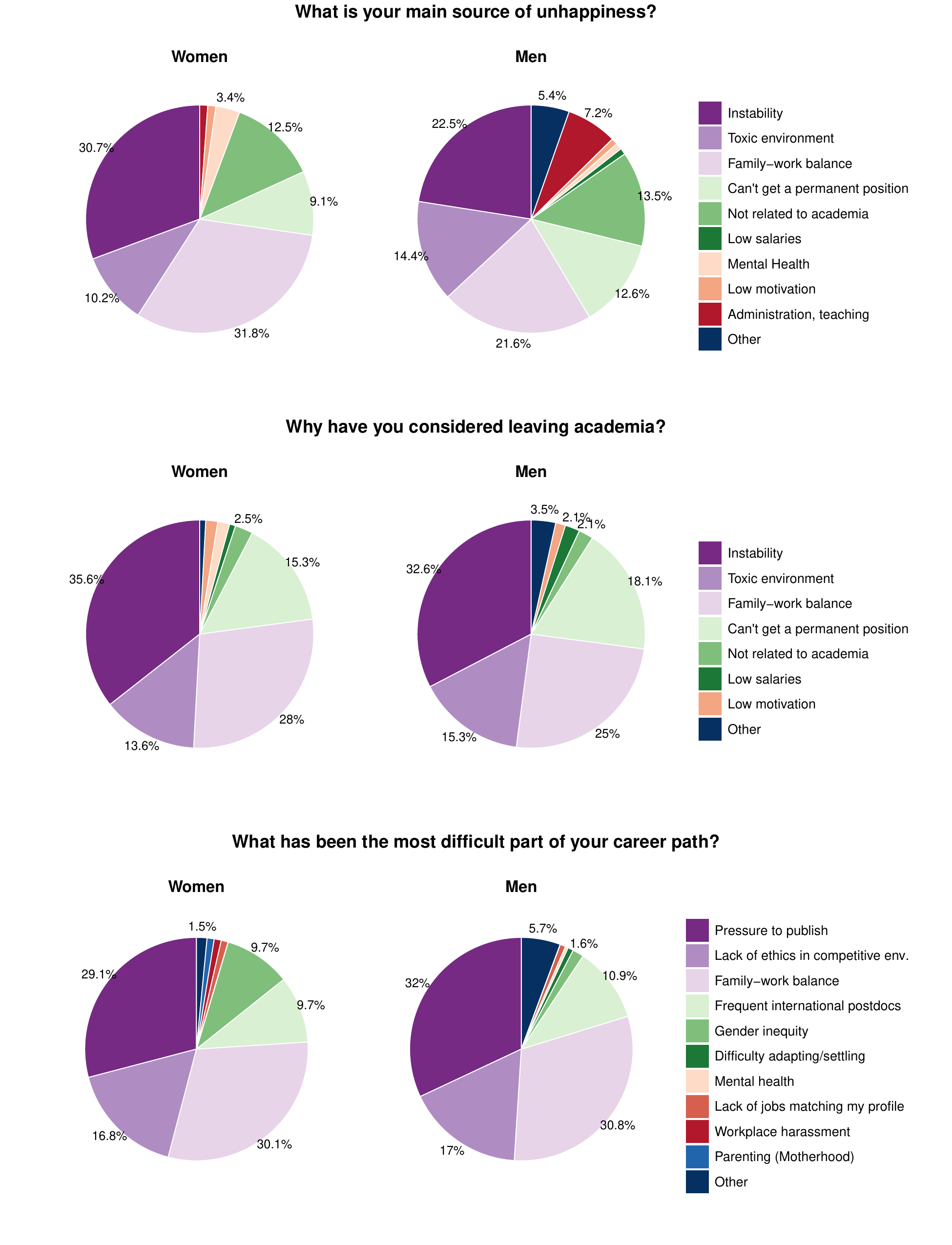}

\caption{Pie charts summarizing survey responses by gender. Each panel displays the distribution of responses to the question indicated in the header. In all panels, the color legend is shown on the right-hand side. The left and right charts correspond to women and men, respectively.}
\label{WhyNotHappy}
\end{figure}

Participants who answered “Sometimes” or “No” to the happiness index question (45\% of the total sample) were asked to indicate the reasons for their unhappiness. Several predefined options were provided, including “Instability”, “Toxic environment”, “Family–work balance”, “Can't get a permanent position”, “Not related to academia”, and “Other”, where participants could specify their own reasons. Due to the repetition of similar answers within the “Other” category, these responses were also grouped and represented in the pie chart shown in Fig.~\ref{WhyNotHappy}. The most frequently mentioned additional reasons included “Low salaries”, “Mental health” , “Low motivation”, and “Administration or teaching duties”. 

Figure~\ref{WhyNotHappy} shows the distribution of responses to this question by gender.  Among women, the main sources of unhappiness are “Family–work balance", followed by “Instability”, “Not related to academia”, and “Toxic environment”. In contrast, men report “Instability” as the most frequent reason, followed by “Family–work balance”, “Toxic environment”, and “Not related to academia”.

%
%
 
We followed up to the question {\it Have you  considered leaving academia?}, people that replied “Yes” ( $\sim$54\%, see  Fig.~\ref{IndexHappy}), were asked  {\it Why have you considered leaving academia?}, approximately 90\% of all replies correspond to “Instability”, “Family–work balance”, and “Difficulty obtaining a permanent position”, in that order, for both genders (see  Figure ~\ref{WhyNotHappy}).

Continuing with the assessment of mental health, participants were also asked, {\it What has been the most difficult part of your career path?} The predefined response options and pie charts are shown in Figure ~\ref{WhyNotHappy}. Overall, both women and men highlighted similar challenges, with “Family–work balance”, “Pressure to publish”, and “Lack of ethics in a competitive environment” among the most frequently cited. However, the relative importance of these factors differed slightly: women most often selected “Family–work balance” as their primary difficulty, whereas men emphasized “Pressure to publish”. 

Among the non-binary (and prefer-not-to-answer) gender group, the overall concerns are very similar: “Instability” is the main reason for unhappiness and for considering leaving academia, while “Pressure to publish” is reported as the most challenging aspect of their career path.


\section{Childcare services}\label{Childcare}

Balancing family responsibilities with professional demands remains a persistent challenge across the scientific community, particularly for researchers with young children. This concern was clearly reflected in the survey responses from the Spanish astronomical community. To better understand this issue, we examined the need for childcare support during the biennial SEA scientific meetings.

As a first step, all surveyed researchers were asked: {\it Do you support the use of SEA funds to provide childcare services during the SEA meetings? }An overwhelming 90\% responded positively, with approximately 57\% of those affirmative responses coming from men, matching the gender distribution of the survey sample (see Fig.~\ref{Childcare}).

Among all participants, 51 researchers indicated that they would require childcare services during the meetings, with a distribution of 55\% women and 45\% men. This corresponds to 32\% of all participating women (28 out of 87) and 18\% of all participating men (23 out of 126). The relationship between this subgroup and two other variables, their index of happiness and whether they had considered leaving academia, is also shown in Fig.~\ref{Childcare}.

Within this subsample, 58\% reported a positive happiness index across all professional categories. Interestingly, all male tenured researchers indicated being happy. Likewise, none of the tenured women reported being unhappy; all answered either “Yes” or “Sometimes”.
However, among those who did not report being happy, 66.6\% were women. Furthermore, approximately 59\% of women in this group stated that they had considered leaving academia. Alarmingly, among female postdoctoral researchers, this proportion rises to about 83\%.

\begin{figure*}[h]
\centering

\includegraphics[width=1.0\textwidth]{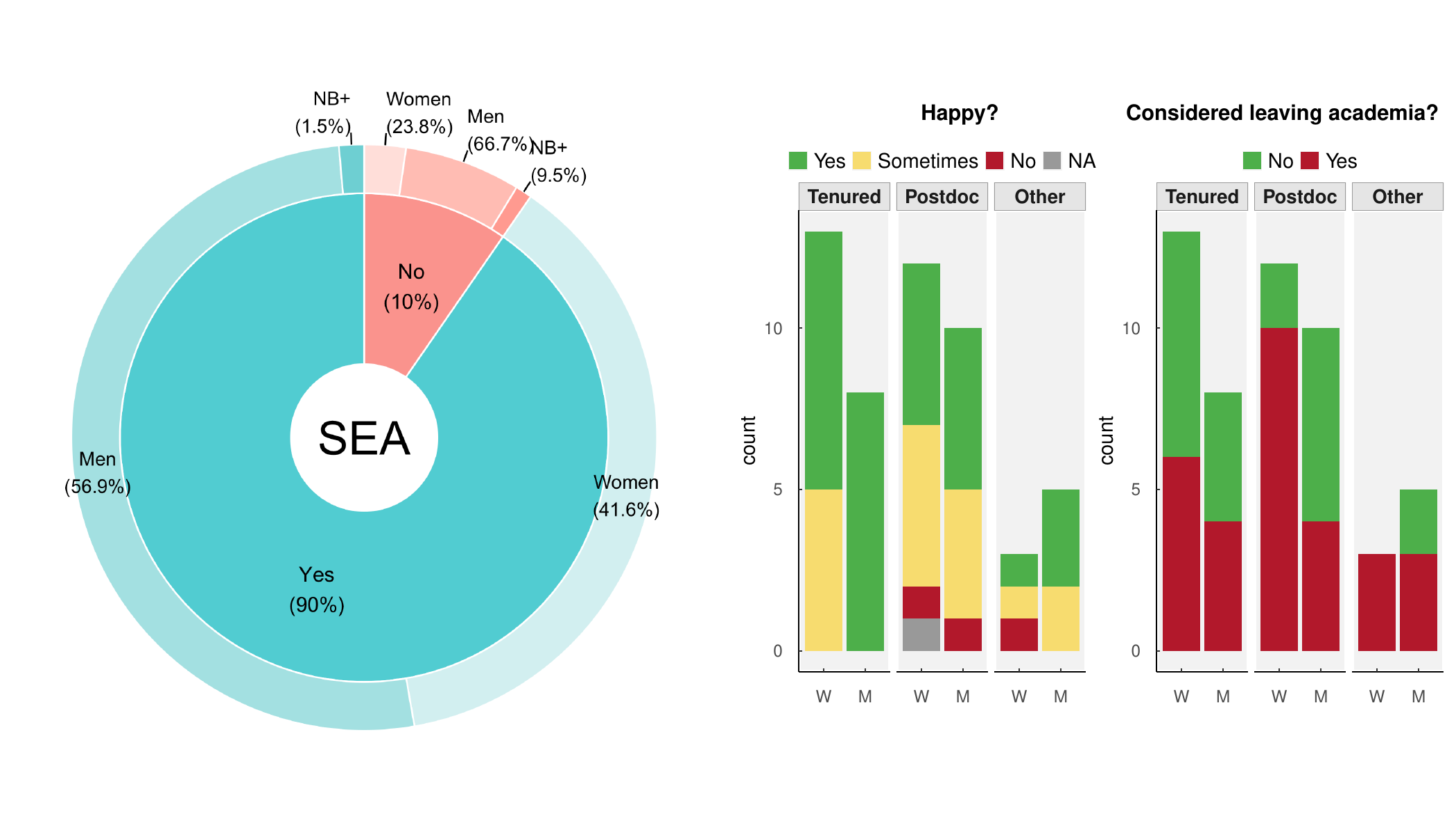}
\caption{Support for childcare services and well-being indicators for researchers requiring childcare. Left: Pie chart showing responses to the question {\it Do you support the use of SEA funds to provide childcare services during the SEA meetings?}, with gender distribution indicated in the outer ring. Right: For researchers who reported needing childcare services during SEA meetings, plots showing gender distribution, happiness index, and whether they have considered leaving academia.}\label{Childcare}
\end{figure*}

\section{Career Instability}\label{Instability}

\begin{figure}[h]
\centering
\includegraphics[width=0.7\textwidth]{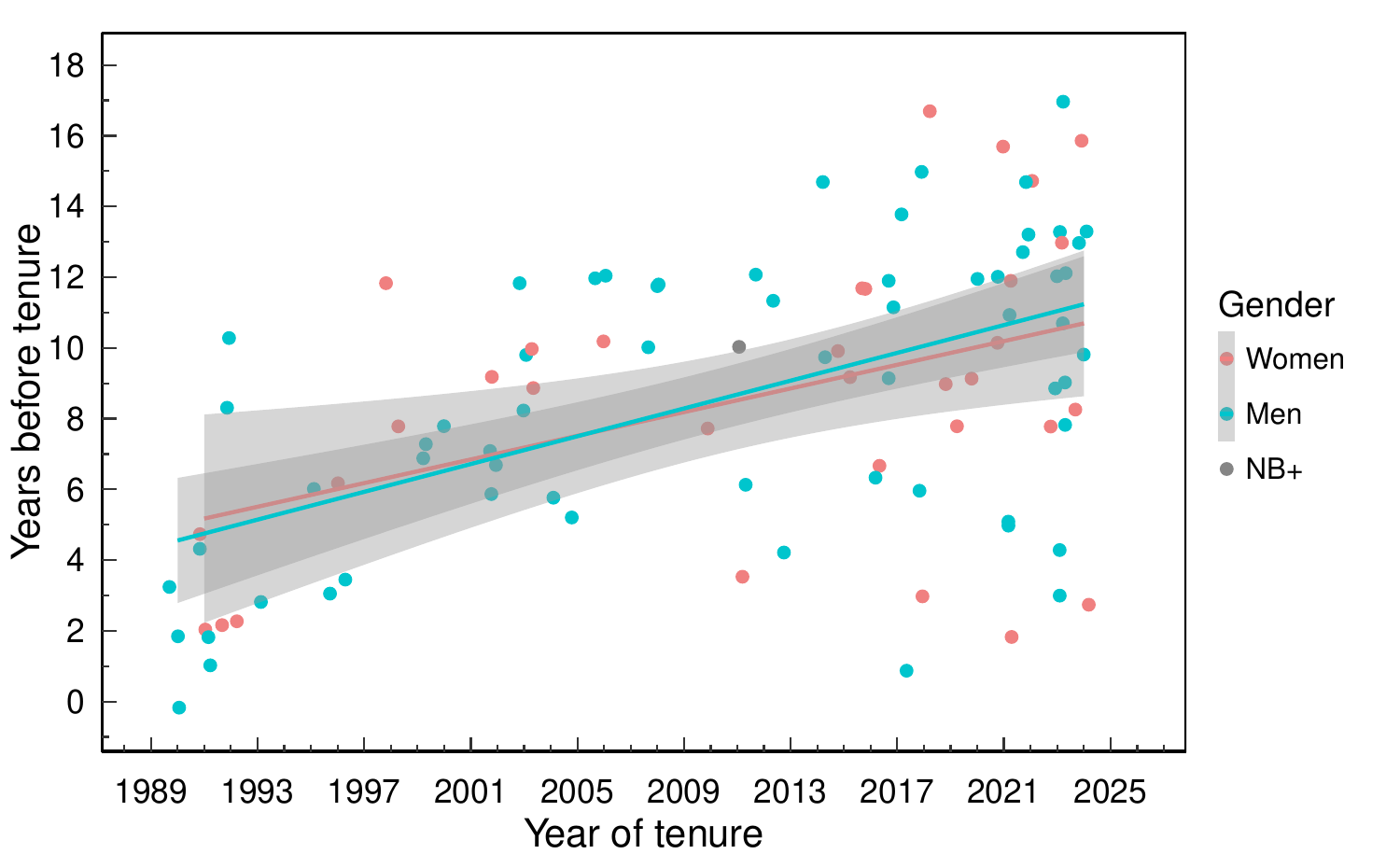}
\caption{Relationship between the year of tenure and the number of years spent as a postdoctoral researcher prior to obtaining tenure, with points color-coded by gender.}\label{YearsPostdocVsTenure}
\end{figure}

Another recurrent source of unhappiness identified in the survey responses relates to instability. To examine this issue in greater detail, we asked all tenured researchers and professors (106 participants) to indicate the year in which they obtained tenure and the number of years they spent as postdoctoral researchers prior to achieving it. The results are shown in Figure~\ref{YearsPostdocVsTenure}. For clarity, a jitter function was applied to both axes to slightly offset overlapping data points.

As illustrated in the figure, before 1993 it took a median of approximately three years after obtaining the PhD to secure a tenured position. However, this period has increased dramatically over time. By 2025, the median time to tenure has risen to around eleven years. When grouped  by gender, Figure~\ref{YearsPostdocVsTenure} shows no significant difference, indicating that both men and women are similarly affected by this growing delay.

In comparison with other countries, this extended postdoctoral period appears consistent with broader international trends. Studies in multiple countries report comparable or even longer pathways to stable employment in academia, particularly in research-intensive fields \cite{KwiekSzymula2024}. Nevertheless, the progressive lengthening of the postdoctoral stage in Spain underscores the increasing difficulty of achieving career stability, a factor that likely contributes substantially to the widespread sense of uncertainty and dissatisfaction reflected in the survey.

 \section{Discussion}\label{Discussion}

Our results highlight several structural and personal factors influencing the well-being of researchers within the Spanish astronomical community, with career instability emerging as one of the most recurrent sources of dissatisfaction. The steady increase in the time required to obtain a tenured position, from a median of three years before 1993 to approximately eleven years by 2025, reflects a broader international trend of prolonged postdoctoral employment and delayed career stabilization. Similar findings have been reported across Europe and the United States, where researchers face increasingly competitive environments, short-term contracts, and limited long-term opportunities \cite{SanzMenendez2013, White2022}. 
Alarmingly, recent findings indicate that nearly half of researchers leave science within a decade \cite{Naddaf2024}, highlighting how systemic instability not only undermines individual well-being but also threatens the retention and diversity of the scientific workforce.

The absence of major gender differences in our data suggests that this instability affects the entire community. However, its interaction with other factors, such as caregiving responsibilities, is likely to place additional strain on women at specific career stages, as observed in the right panel of Fig. \ref{Childcare}. Indeed, numerous studies report that women tend to leave academia earlier and in greater proportions than men \cite{Alper1993, Blickenstaff2005, DeutschYao2014, Goulden2011, ShawStanton2012}.



The relationship between work–family balance and well-being also emerged as a central concern. Responses from both women and men frequently mentioned family–work imbalance, instability, and pressure to publish as dominant stressors. The data further indicate that among women postdoctoral researchers, more than 80\% have considered leaving academia, highlighting the cumulative impact of instability and caregiving responsibilities during the most vulnerable career stages.

Our analysis of childcare needs reinforces this interpretation. An overwhelming 90\% of participants supported the allocation of SEA funds to provide childcare services during biennial scientific meetings, underscoring the community’s recognition of this barrier to participation. Fifty-one participants indicated that they would personally require such services, with nearly equal representation of women and men. Despite this parity, women who expressed a need for childcare reported lower happiness indices and a greater tendency to consider leaving academia, suggesting that the provision of institutional support for caregiving may directly enhance retention and well-being across genders \cite{Calisi2018}.

To date, the SEA scientific meetings do not provide childcare services. However, the Women and Astronomy Commission is actively advocating for their implementation. We hope that the publication of these results will contribute to raising awareness of this issue and support the adoption of childcare facilities in future SEA meetings, fostering a more inclusive and equitable environment for all members of the community.
Moreover, such progress has the potential to inspire other national and international scientific meetings to adopt similar initiatives, contributing to broader efforts to improve equity and retention across the global research community.

Although our analysis focuses on the SEA community, it is important to emphasize that the lack of childcare provision at scientific meetings is not a problem unique to Spain. Similar concerns have been raised across the international research community \cite{Calisi2018}, where the absence of family-support policies continues to limit full participation, particularly for early-career researchers and caregivers.

Finally, although this study focuses on members of the Spanish Astronomical Society, the structural challenges identified here, such as  career instability, pressure to publish, and difficulties in reconciling academic careers with family responsibilities, are widely reported across the international research community. Similar trends have been documented in Europe \cite{Boissier12,SanzMenendez2013} and North America \cite{White2022}, suggesting that the patterns observed here are not unique to Spain, but instead reflect a broader, systemic crisis in well-being and mental health across astronomy and related disciplines. While national contexts may differ in funding structures, academic culture, and career pathways, our findings add to a growing body of evidence highlighting shared challenges and may therefore be informative for other countries and scientific societies seeking to improve researcher well-being and retention.


\section{Conclusion}\label{Conclusion}

The findings of this study illustrate that career instability, work–family imbalance, and the pressure to publish are key drivers of unhappiness within the SEA community, consistent with broader trends observed across the global research landscape. The growing delay between PhD completion and the attainment of a permanent position signals an urgent need to reassess academic career structures and support mechanisms.

Promoting policies that facilitate career security, flexible mobility, and accessible childcare services would directly address several of the most critical sources of dissatisfaction identified in our survey. The strong support within the astronomy community for childcare initiatives at SEA meetings demonstrates a collective commitment to fostering an inclusive and family-friendly environment.

Ultimately, improving well-being in academia requires coordinated efforts between institutions, funding agencies, and professional societies to ensure that scientific excellence is compatible with personal stability and work–life balance. These findings provide evidence-based motivation for such reforms and highlight the importance of sustained monitoring of the professional and personal conditions shaping researchers’ lives.
\backmatter

%
%

\bmhead{Acknowledgements}

We are deeply grateful to Dr. Ali Lara, social psychologist, for his guidance in formulating the happiness-related questions.

\section*{Declarations}

\begin{itemize}
\item Funding: MALL  acknowledges grants  PID2024-155875OB-I00 funded by MICIU/AEI/10.13039/501100011033/FEDER, EU  and RYC2020-029354-I funded by MICIU/AEI/10.13039/501100011033 by “ESF Investing in your future”, by “ESF+” .
AVG acknowledges support from the Spanish grant PID2022-138560NB-I00, funded by MCIN/AEI/10.13039/501100011033/FEDER, EU.
IGB is supported by the Programa Atracci\'on de Talento Investigador ``C\'esar Nombela'' via grant 2023-T1/TEC-29030 funded by the Community of Madrid.

\item Conflict of interest/Competing interests: 
The authors declare no competing interests

\item Authors contribution:  MALL proposed to the design of the survey questions, performed the data analysis, generated all figures, and drafted the manuscript. IR, AVG, SRB, ARE, SRB, IGB, BAG, MRB,  NBI, IPC, NO and SB provided sustancial feedback on the manuscript and survey questions. 

\end{itemize}

\noindent

\bibliography{SEA_WellBeing}

\end{document}